%% file: main.tex
\documentclass{article}

\usepackage{microtype}
\usepackage{graphicx}
\usepackage{subfigure}
\usepackage{booktabs} 
\usepackage{enumitem}

\usepackage{hyperref}

\usepackage[accepted]{icml2023_deployableGenAI}

\usepackage{amsmath}
\usepackage{amssymb}
\usepackage{mathtools}
\usepackage{amsthm}
\usepackage{pgfplots} 
\usepackage{caption}
\pgfplotsset{compat=1.18}

\usepackage{tikz}

\usepackage[capitalize,noabbrev]{cleveref}

\theoremstyle{plain}

\theoremstyle{definition}

\theoremstyle{remark}

\usepackage[textsize=tiny]{todonotes}

\usepackage{tablefootnote}

\icmltitlerunning{Data-Driven Generative Audio AI}

\begin{document}

\twocolumn[
\icmltitle{A Demand-Driven Perspective on Generative Audio AI}

\icmlsetsymbol{equal}{*}

\begin{icmlauthorlist}
\icmlauthor{Sangshin Oh}{equal,comp}
\icmlauthor{Minsung Kang}{equal,comp}
\icmlauthor{Hyeongi Moon}{comp}
\icmlauthor{Keunwoo Choi}{comp}
\icmlauthor{Ben Sangbae Chon}{comp}
\end{icmlauthorlist}

\icmlaffiliation{comp}{Gaudio Lab, Inc., Seoul, South Korea}

\icmlcorrespondingauthor{Ben Sangbae Chon}{bc@gaudiolab.com}

\icmlkeywords{Audio AI, ICML}

\vskip 0.3in
]



\printAffiliationsAndNotice{\icmlEqualContribution} 

\begin{abstract}
To achieve successful deployment of AI research, it is crucial to understand the demands of the industry. In this paper, we present the results of a survey conducted with professional audio engineers, in order to determine research priorities and define various research tasks. We also summarize the current challenges in audio quality and controllability based on the survey. Our analysis emphasizes that the availability of datasets is currently the main bottleneck for achieving high-quality audio generation. Finally, we suggest potential solutions for some revealed issues with empirical evidence.

\end{abstract}

\input{main_text}
\bibliography{example_paper}
\bibliographystyle{icml2023}

\input{appendix}


\end{document}

%% file: main_text.tex
\section{Introduction}
\label{submission}
The use of audio generative models has the potential to significantly impact a variety of industries.  Although essential, the process of creating foley effects is often tedious, non-reproducible, and lacks scalability. Moreover, the utilization of pre-recorded sounds is not conducive to real-time or interactive applications, rendering it inadequate for fields like gaming, metaverse, or any domain requiring the simulation of lifelike environments. The advent of generative audio AI offers a promising solution to address these limitations, significantly impacting areas like film production, gaming, social platforms, and more.

Audio synthesis research has a long history \cite{dudley1955fundamentals, chowning1973synthesis}, but we will focus on the data-driven approaches as they are the recent pioneers with huge potential.
The current generative audio AI is still in its early stages, necessitating further advancements in various aspects. We present this paper to provide a demand-driven perspective on task definitions, challenges, and potential solutions within audio generation. Specifically, our focus is on general audio, excluding speech and music. 

The key contributions of this paper include:

\begin{itemize}[leftmargin=*, topsep=-0.1em]
  \setlength\itemsep{-0.15em}
    \item A survey with individuals working in movie sound productions to share insights into the industry-side demands.
    \item Detailed definitions and review of distinct tasks in audio generation regarding input types and conditions.
    \item A summary of the related challenges towards industrial demands and a proposal on potential solutions supported by empirical evidence, including a method with which we achieved 2nd place in the foley synthesis challenges at DCASE 2023.
\end{itemize}

\section{Demands from Industry}
\label{sec:survey}
To gather insights regarding the impact of audio generative models on the industry, we first interviewed two professionals from the field of movie sound production. They highlighted that their role extends beyond that of sound technicians, as they contribute to the artistic dimension of creating immersive and captivating sound experiences.
Despite the inevitable laborious nature of foley and sound effect recording, they are compelled to record new sounds since existing sounds are hardly reusable. While they have a vast library of previous sound stems, there is effectively no efficient method at hand for searching and finding suitable sounds. Even if they find a suitable sound, they have to spend time on editing the time synchronization and sound tone.

Based on this knowledge, we conducted a survey involving 18 individuals working in movie sound production, addressing the topic of AI audio generation. We first presented them with some examples of AI image generation applications and a demo page\footnote{\url{https://audioldm.github.io/}} of a recent text-to-audio model~\cite{liu2023audioldm}.
We then asked three following primary questions with multiple-choice options.

\textbf{Q1.} \textit{What are the major challenges faced in foley recording?} The most frequently selected option for this question was the time synchronization problem. Following that, respondents expressed the importance of audio quality and consistency in tone with the synchronous recording. 
In the additional comments, respondents emphasized again that for foley sound, audio quality, synchronization with the scene, and consistency in tone with other sound sources are crucial -- to the point that without a good synchronization, some might only consider using AI-generation for ambient sounds.
This indicates that relying solely on text-based conditioning may not be sufficient for a majority of use-cases. 

\begin{figure}[t]
\centering
\begin{tikzpicture}
\pgfplotsset{%
    width=.46\textwidth,
    height=0.29\textwidth
}
\begin{axis}[
    xbar,
    tick align=outside,
    enlargelimits=0,
    enlarge y limits=0.115,
    nodes near coords,
    nodes near coords align=horizontal,
    nodes near coords style={
        anchor=west,
    },
    xlabel={Number of votes},
    xmin=0,
    xmax=15,
    xtick={0,1,2,3,4,5,6,7,8,9,10,11,12,13,14,15},
    ytick={data},
    yticklabels={\textit{sync},\textit{speed},\textit{copyright},\textit{text},\textit{edit},\textit{creativity},\textit{quality}},
    xticklabels={0,,2,,4,,6,,8,,10,,12,,14},
    point meta=explicit symbolic,
]
\addplot[draw=gray,fill=gray!15] coordinates {
    (2,1)[11.1\%]
    (3,2)[16.7\%]
    (3,3)[16.7\%]
    (6,4)[33.3\%]
    (11,5)[61.1\%]
    (11,6)[61.1\%]
    (12,7)[66.7\%]
};
\end{axis}
\end{tikzpicture}%
\caption{Answers of Q2: \textit{What is the limitation(s) of the current text-conditioned audio generation as a product?}}
\label{fig:difficulties}
\end{figure}
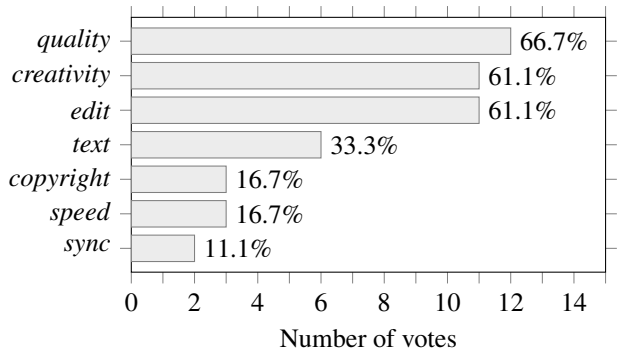

\textbf{Q2.} \textit{What is the limitation(s) of the current text-conditioned audio generation as a product?}   
The survey result is plotted in Figure \ref{fig:difficulties}. In this question, it was found that audio quality presents the most significant challenge for practical usage. According to their comments, the concerns about \textit{quality} encompass other aspects such as low fidelity, low sampling rate, roughness, and other related factors. A majority of respondents expressed complaints regarding the sample rate. It is noteworthy that while the industry requires full-band signals at 48kHz or higher, most of the current systems still operate within the 16kHz-24kHz range \cite{kreuk2022audiogen, huang2023make, liu2023audioldm}. 
For \textit{creativity}, which was the second most frequently chosen category, it refers to the generation of new sounds that fulfill artistic intentions, e.g., creating ``the sound of a lightsaber in Star Wars." The terms such as \textit{edit} and \textit{text}, which received the third and fourth highest numbers of votes, indicate the problems of controllability.

\begin{figure}[t]
\centering
\begin{tikzpicture}
\pgfplotsset{%
    width=.47\textwidth,
    height=.24\textwidth
}
\begin{axis} [
    xbar,
    tick align=outside,
    enlargelimits=0,
    enlarge y limits=0.15,
    nodes near coords,
    nodes near coords align=horizontal,
    nodes near coords style={
        anchor=west,
    },
    xlabel={Number of votes},
    xmin=0,
    xmax=15,
    xtick={0,1,2,3,4,5,6,7,8,9,10,11,12,13,14,15},
    ytick=data,
    yticklabels={\textit{image},\textit{consistn.},\textit{interp.},\textit{ref. aud.},\textit{video}},
    xticklabels={0,,2,,4,,6,,8,,10,,12,,14},
    point meta=explicit symbolic,
]

\addplot[draw=gray,fill=gray!15] coordinates {
    (5,1)[27.8\%]
    (7,2)[38.9\%]
    (7,3)[38.9\%]
    (10,4)[55.6\%]
    (12,5)[66.7\%]
};

\end{axis}
\end{tikzpicture}%
\caption{Answers of Q3: \textit{How would you like to condition the audio generation by?}}
\label{fig:condition}
\end{figure}
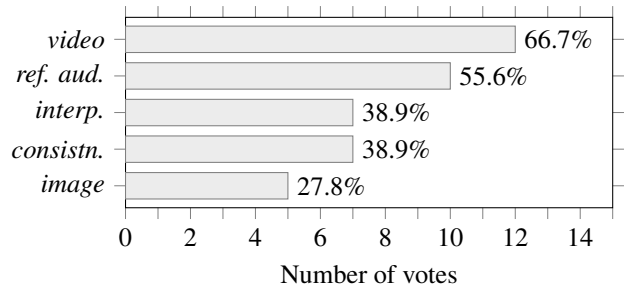

\textbf{Q3.} \textit{How would you like to condition the audio generation?} As in Figure \ref{fig:condition}, the most frequently chosen option is the utilization of video for time synchronization and achieving an appropriate sound tone. More than half of the respondents were interested in generating similar sounds to reference audio samples. The third and fourth popular options, namely \textit{interp.} and \textit{consistn.}, are related to refining the generated audio based on reference audio samples. The respondents seemed to show their hope for a more efficient workflow in Q3, in contrast to showing their expectations in Q2.   

This survey result presents important remarks on generative audio research. First, texts and videos are complementary to each other towards a more complete generative audio system. Second, sound and event synchronization is an important topic that deserves more attention. Third, although it is somewhat deviated from our topic, high-quality audio indexing, search, and separation may be also a solution for some of the problems generative audio AI aims to solve. 
Based on this understanding, we delve into the current state and challenges of the audio generation field in the following sections.

\section{Task Definitions}
In a recent proposal paper on foley sound synthesis challenge \cite{choi2022proposal}, the audio generative AI task is specified based on the input and output types. The authors outline three distinct input types: i) category index, ii) text description, and iii) videos. While the categorization of output types is not explicitly stated, it can be inferred as follows: i) individual foley sounds representing a single event, ii) a combination of multiple events and/or ambient sounds, and iii) a comprehensive soundtrack comprising foley sounds ambient elements, and spatially enhanced mixing. We will focus on the input types since the determination of output types is primarily governed by technical feasibility, allowing a limited scope with the current technology.

\subsection{Input Types}

First, a category index, that indicates a single type of audio event, would be the simplest form of input type for a sound synthesis system. This was adopted in some previous works \cite{ghose2020autofoley, ghose2022foleygan} and this year's DCASE Task 7 \cite{choi2023foley}. Solutions with this approach would improve foley recording processes for some popular categories such as dog barks, door slams, or footsteps.

The second type would be text descriptions as employed in recent research \cite{kreuk2022audiogen, yang2023diffsound, liu2023audioldm, huang2023make}, relying on audio caption datasets.
%
There are several promising aspects associated with this text-to-audio approach. i) Extensive research has already been conducted on text-to-X generation (e.g., text-to-image generation studies \cite{ramesh2021zero, saharia2022photorealistic, rombach2022high}), which simplifies its adaptation for audio generation purposes. ii) The familiarity of users with UI/UX utilizing text inputs further supports the feasibility of this approach. 
However, there are difficulties as well. i) Compared to text-image pairs, there is a scarcity of text-audio pairs available for training models \cite{huang2023make}.
For example, the number of items of AudioCaps \cite{kim-etal-2019-audiocaps}, the largest audio captioning dataset, is 0.013\% of (or 7561 times smaller than) that of LAION-400M, an text-image pair dataset~\cite{schuhmann2021laion}. 
ii) Text input has limitations in providing highly detailed descriptions at a professional level, as audio engineers rely on precise controls like knobs and sliders to make fine adjustments to the sound (e.g., equalizers).

Third, video input types have pros and cons. Unlike the previous input types, videos may provide the exact timings of events \cite{zhou2018visual, ghose2022foleygan, cui2023varietysound}. As its importance was discussed in Section~2, there is a huge potential for improving the workflow of video creation in this scenario by efficient time synchronization. However, the video itself does not provide complete information because it is common that not everything visible should sound, as well as not everything that sounds is visible. 
Additionally, there are deliberate artistic intentions involved in video creation such as muting/exaggerating certain sounds. These artistic decisions may vary significantly. Therefore, when developing video-to-sound generation methods, the ability to edit and manipulate the generated audio becomes crucial, just as it is important for text-based generation approaches as we will discuss in the following section.

\subsection{Conditioning}
Conditioning can be viewed as a form of input in a broader sense and is deeply related to controllability and editability. AudioLDM pioneered sound editing through text-based approaches~\cite{liu2023audioldm}, and we believe that this direction of research will continue toward more diverse, intuitive, and fine-grained conditioning. For example, users may want to control factors such as sound bandwidth, F0 contours, temporal and spectral envelopes, etc.   Our exploration of these product development considerations will continue in the following sections.

\section{Challenges}
\label{sec:challenges}

\subsection{Dataset Improvement for Audio Quality}
Recently, there have been some generative AI products successfully deployed on language and image~\cite{touvron2023llama, chowdhery2022palm, openai2023gpt4, ramesh2021zero}. However, the current state of audio generation research does not seem mature enough to be adopted into professional sound production. As audio quality was the most prominent issue as in Figure~\ref{fig:difficulties}, we focus on the issues and potential solutions on datasets to improve the generated audio quality in this section.

\input{dstable}

First of all, the current data scarcity deteriorates the model training and resulting audio quality. Compared to image generation datasets that go beyond a few billion pairs~\cite{ramesh2022hierarchical}, there are much less text-paired audio data available \cite{kreuk2022audiogen, huang2023make}. Moreover, most of such paired datasets are \textit{weakly labeled}, i.e. their labels or captions lack time resolution. This is problematic as it is common practice to slice audio signals for ease of training and memory-related issues.
Since the text in the pairs depicts audio coarsely in the time axis, there should be potential risks of mismatching when the audio signal is sliced into smaller segments for some practical reasons.
Augmentation method \cite{kreuk2022audiogen, huang2023make} or using a contrastive embedding network \cite{liu2023audioldm} can help this, but not as an absolute treatment.

The characteristics of the audio itself even exacerbates the problem. It is a difficult problem to separate foreground and background audio sources, and obtaining isolated audio recording would remain to be costly. The spatial characteristics of the recording environment often have negative affects to the recording quality. Altogether, there are many factors that make it tricky to create a studio-quality audio dataset. We listed available audio datasets in Table~\ref{tab:dataset}. Since the largest datasets in the list are collected or curated from crowd-sourced audio \cite{font2013freesound} or video \cite{gemmeke2017audio, chen2020vggsound}, their recording conditions may vary and are usually not good. Thus, the samples from those datasets often suffer from severe background noises, low recording bandwidth / bit rate, and various types of distortion. \textit{Clean} datasets are limited to several commercial sound effect libraries.

To this trade-off problem of \textit{more} data vs. \textit{clean} data, we propose a solution called \textit{quality-aware training} (QAT). This can be simply done by prompting, i.e., appending dataset labels indicating the quality of the dataset in the text input. QAT enables to utilize a broader range of datasets. During the training phase, a model can learn from both \textit{clean} and \textit{noisy} datasets with quality labels. As a result, the model would learn not only the concepts of different audio events but also their audio quality; i.e., the model would have \textit{compositionality} of audio events and audio quality. During the inference phase, we can force the model to generate clean signals by conditioning the model, i.e., by appending `clean' labels to the text input. 
This enabled us to use all data pairs regardless of their quality without deteriorating their output quality. 
In our experience, this approach let us control the audio quality, reverberation, signal bandwidth, and audio event independently and achieve 2nd place in the recent foley synthesis challenge at DCASE 2023 \cite{ChonGLI2023, kang2023falle}. Details about experiments are provided in Appendix B.

\subsection{Methodological Improvement for Controllability}
Controllability was another major concern in our survey, as the audio engineers have specific intent about how the generated output should sound. Audio generation may take a long time, hence it is crucial for deployable audio AI systems to have effective controllability 

Classifier-free guidance is a widely adopted solution for the problem across diffusion-based and Transformer-based generative models. At the cost of sample qualities by extrapolating intermediate features or logits, it introduces diversity, which would make exploration easier for the users of generative audio AI systems. Most of the recent text-to-audio generation research adopted this technique \cite{kreuk2022audiogen, liu2023audioldm, huang2023make}.

Controllability can be also attained by introducing new features or new modalities, for example, a reference audio or a conditioning video as in Figure~\ref{fig:condition}. As AudioLDM demonstrated audio manipulation without fine-tuning \cite{liu2023audioldm}, we believe text-guided audio-to-audio generation is a compelling research direction towards deployable generative audio AI. Video-based foley generation has been less popular, but it would be an interesting direction for future research along with the existing research~\cite{zhou2018visual, ghose2020autofoley, ghose2022foleygan}. Finally, conventional signal features such as F0 contour or envelopes can be a great user interface for experienced audio engineers. As those features are easy to extract from audio signals, it is plausible to use them as one of the inputs during the training phase, then build a user interface that allows control of the generated output by modifying the features. 

\section{Conclusion}
In this paper, we presented a survey conducted with sound engineers in the movie industry. Based on the survey results, we have provided task definitions for audio generation research and identified related research challenges. Our objective was to bridge the gap between current research and industry practices, offering potential solutions to address the challenges of audio quality and controllability.

Surprisingly, there are limited opportunities for researchers to gain insights from the industry side. We believe that this work serves as a valuable starting point for understanding the difficulties faced by both researchers and potential users, ultimately aligning our efforts to solve the real-world problems.

While our perspective focuses on the movie industry, it is important to acknowledge that neighboring industries may face different challenges with varying priorities. For example, the demand for real-time generation systems may be stronger in the virtual reality or gaming industry, while the standards for audio quality or artistic intent may be lower for non-professional movie creation platforms such as YouTube. We hope that our work represents a meaningful step towards comprehending the diverse demands placed on generative audio AI and its diverse applications.


%% file: dstable.tex
\begin{table}[t]
\centering
\resizebox{\columnwidth}{!}{%
\begin{tabular}{l c r r c c c}
\toprule
\textbf{Name} & \textbf{AQ}
    & \multicolumn{2}{c}{\textbf{Dataset Size}}
    & \multicolumn{3}{c}{\textbf{Modality}} \\
\cmidrule(r){3-4} \cmidrule(l){5-7}
&  
& \multicolumn{1}{c}{\textbf{Dura.}} & \multicolumn{1}{c}{\textbf{N. Files}}
& \multicolumn{1}{c}{\textbf{Lb}} & \multicolumn{1}{c}{\textbf{Cp}} &\multicolumn{1}{c}{\textbf{Vd}} \\
\midrule
\textbf{Audioset} \\
\parbox[t]{3cm}{AudioSet \\ \cite{gemmeke2017audio}}
    & \textit{noisy}    & 5420 \textit{h}   & 1,951,460   & \checkmark  &  & \checkmark \\
\parbox[t]{3cm}{AudioCaps \\ \cite{kim-etal-2019-audiocaps}}
    & \textit{noisy}    & 144.9 \textit{h}  & 52,904       & \checkmark    & \checkmark   & \checkmark \\
\midrule
\textbf{Freesound} \\
\parbox[t]{3cm}{Freesound \\ \cite{font2013freesound} }
    & \textit{noisy}    & 3003 \textit{h}   & 515,581   & & $\triangle$   &   \\
\parbox[t]{3cm}{UrbanSound8K \\ \cite{salamon2014dataset}}
    & \textit{noisy}     & 8.75 \textit{h}   & 8,732     & \checkmark & & \\
\parbox[t]{3cm}{ESC-50 \\ \cite{piczak2015esc}}
    & \textit{noisy}    & 2.78 \textit{h}   & 2,000 & \checkmark &  &   \\
\parbox[t]{3cm}{Clotho\\ \cite{drossos2020clotho}}
    & \textit{noisy}    & 37.0 \textit{h}   & 5,929 &   & \checkmark &\\
\parbox[t]{3cm}{FSD50K\\ \cite{fonseca2021fsd50k}}
    & \textit{noisy}    & 108.3 \textit{h}  & 51,197    & \checkmark &  &   \\
\midrule
\textbf{Others} \\
\parbox[t]{3cm}{VGG Sound \\ \cite{chen2020vggsound} }
    & \textit{noisy}    & 550 \textit{h}   &  $\approx$ 200,000     & \checkmark &  & \checkmark\\
BBC sound effects \footnotemark
    & \textit{clean}    & 463.5 \textit{h}    & 15,973         &                            & \checkmark    & \\
Epidemic Sound effects \footnotemark
    & \textit{clean}     & 220.4 \textit{h}  & 75,645    & \checkmark & & \\
Free To Use Sounds \footnotemark
    & \textit{noisy}    & 175.7 \textit{h}  & 6,370     &   & \checkmark &  \\
Sonnis Game Effects \footnotemark
    & \textit{clean}    & 84.6 \textit{h}   & 5,049     &   & $\triangle$   & \\
WeSoundEffects \footnotemark
    & \textit{clean}    & 12.0 \textit{h}   &   488     &   &   $\triangle$     &   \\
Odeon Sound Effects \footnotemark
    & \textit{clean}    & 19.5 \textit{h}   & 4,420     &   & $\triangle$  &   \\
\bottomrule
\end{tabular}%
}
\caption{
A list of audio datasets. AQ: audio quality, Dura.: duration, N. Files: number of files. Modality columns refer to the existence of labels, captions, and videos, respectively. 
\textit{Clean} recording: Audio is recorded in well-treated environments and mastered for professional content production. \textit{Noisy}: dataset contains environmental noises or interference signals.
$\triangle$: Textual information included, not necessarily captions.
This table is partially from \cite{kreuk2022audiogen} and \cite{wu2023large}
}
\label{tab:dataset}
\end{table}

\addtocounter{footnote}{-5}
\footnotetext{\small \url{https://sound-effects.bbcrewind.co.uk}}
\addtocounter{footnote}{1}
\footnotetext{\small \url{https://www.epidemicsound.com/sound-effects/}}
\addtocounter{footnote}{1}
\footnotetext{\small  \url{https://www.freetousesounds.com/all-in-one-bundle/}}
\addtocounter{footnote}{1}
\footnotetext{\small \url{https://sonniss.com/gameaudiogdc}}
\addtocounter{footnote}{1}
\footnotetext{\small  \url{https://wesoundeffects.com/we-sound-effects-bundle-2020/}}
\addtocounter{footnote}{1}
\footnotetext{\small  \url{https://www.paramountmotion.com/odeon-sound-effects}}

%% file: appendix.tex
\newpage
\appendix
\onecolumn

\section{Details of survey in Section \ref{sec:survey}}

\subsection{Exact expression of the options in Figure \ref{fig:difficulties} and Figure \ref{fig:condition}}

\begin{table}[H]
\centering
\begin{tabular}{l l p{0.75\linewidth}}
\toprule

Question & option & exact expression \\

\midrule
Figure \ref{fig:difficulties} & \textit{quality} & Audio quality. \\
& \textit{creativity} & Lack of creativity in fulfilling artistic intentions (e.g. the sound of lightsabers in Star Wars). \\
& \textit{edit} & Detailed audio editing (e.g. I like the footstep sound I've created, but I wish it was a bit lighter). \\
& \textit{text} & Difficult to create the desired sound with just text. \\
& \textit{copyright} & Copyright. \\
& \textit{speed} & Speed of generation. \\
& \textit{sync} & Time synchronization with the scene. \\
\midrule
Figure \ref{fig:condition}  & \textit{video} & Time synchronization and incorporating tone through video. \\
& \textit{ref. aud.} & Create sounds similar to a reference (e.g., ``Create 10 sounds similar to Sound A" or``Make a slightly more light version of Sound A"). \\
& \textit{interp.} & 
Interpolation of two sounds (e.g., ``I need a footstep sound that is a middle ground between Sound A and Sound B"). \\
& \textit{consistn.} & Generate sounds consistent with the reference audio (e.g., ``Create a car sound that matches the 0:00:32 - 0:00:42 segment of this track"). \\
& \textit{image} & Expressing the sensation of sound that is difficult to convey in words through image. \\

\bottomrule
\end{tabular}
\caption{}
\end{table}

\subsection{Results on the other questionnaire}

\begin{figure}[H]
\centering
\begin{tikzpicture}[trim axis left]
\begin{axis} [
    xbar,
    width=0.6\textwidth,
    y=17pt,
    tick align=outside,
    enlargelimits=0,
    enlarge y limits={abs=0.6},
    nodes near coords,
    nodes near coords align=horizontal,
    nodes near coords style={
        anchor=west,
    },
    xlabel={Number of votes},
    xmin=0,
    xmax=12,
    xtick={0,1,2,3,4,5,6,7,8,9,10,11,12,13,14,15,16},
    ytick=data,
    yticklabels={
        \textit{have heard of T},
        \textit{have used T},
        \textit{often use T},
        \textit{have heard of I},
        \textit{have used I},
        \textit{often use I},
    },
    xticklabels={0,,2,,4,,6,,8,,10,,12,,14,,16},
    point meta=explicit symbolic,
]

\addplot[draw=gray,fill=gray!15] coordinates {
    (10,6)[55.6\%]
    (3,5)[16.7\%]
    (0,4)[0\%]
    (9,3)[50.0\%]
    (1,2)[5.6\%]
    (0,1)[0\%]
};

\end{axis}
\end{tikzpicture}%
\captionsetup{width=0.8\textwidth}
\caption{Answers of a multiple-choice question. \vspace{5pt} \\
\textbf{Q}: Have you heard of or used text or image generation AI, such as ChatGPT, Bard, Stable Diffusion or Midjourney? \vspace{3pt} \\
\textbf{A1}: \textit{have heard of T} - Have heard of text generative model \\
\textbf{A2}: \textit{have used T} - Have used text generative model \\
\textbf{A3}: \textit{often use T} - Often use text generative model \\
\textbf{A4}: \textit{have heard of I} - Have heard of image generative model \\
\textbf{A5}: \textit{have used I} - Have used image generative model \\
\textbf{A6}: \textit{often use I} - Often use image generative model \\
        }
\end{figure}

\begin{figure}[H]
\centering
\begin{tikzpicture}[trim axis left]
\begin{axis} [
    xbar,
    width=0.6\textwidth,
    y=17pt,
    tick align=outside,
    enlargelimits=0,
    enlarge y limits={abs=0.6},
    nodes near coords,
    nodes near coords align=horizontal,
    nodes near coords style={
        anchor=west,
    },
    xlabel={Number of votes},
    xmin=0,
    xmax=12,
    xtick={0,1,2,3,4,5,6,7,8,9,10,11,12,13,14,15,16},
    ytick=data,
    yticklabels={
        \textit{time saving},
        \textit{replace rec.},
        \textit{high-quality},
        \textit{copyright-free},
        \textit{ambience},
        \textit{cost saving},
    },
    xticklabels={0,,2,,4,,6,,8,,10,,12,,14,,16},
    point meta=explicit symbolic,
]

\addplot[draw=gray,fill=gray!15] coordinates {
    (10,6)[55.6\%]  
    (9,5)[50.0\%]  
    (9,4)[50.0\%]  
    (3,3)[16.7\%]  
    (1,2)[5.6\%]  
    (1,1)[5.6\%]  
};

\end{axis}
\end{tikzpicture}%
\captionsetup{width=0.8\textwidth}
\caption{Answers of a multiple-choice question. \vspace{5pt} \\
\textbf{Q}: What do you expect for generative audio AI product? \vspace{3pt} \\
\textbf{A1}: \textit{time saving} - Time saving due to the fast generation speed \\
\textbf{A2}: \textit{replace rec.} - Replacing recording or sampling process with generation \\
\textbf{A3}: \textit{high-quality} - Obtaining high-quality, well-aligned audio \\
\textbf{A4}: \textit{copyright-free} - Obtaining copyright-free sources \\
\textbf{A5}: \textit{ambience} - Generating ambient sound \\
\textbf{A6}: \textit{cost saving} - Cost saving for human resources \\
        }
\end{figure}

\begin{figure}[H]
\centering
\begin{tikzpicture}[trim axis left]
\begin{axis} [
    xbar,
    width=0.6\textwidth,
    y=17pt,
    tick align=outside,
    enlargelimits=0,
    enlarge y limits={abs=0.6},
    nodes near coords,
    nodes near coords align=horizontal,
    nodes near coords style={
        anchor=west,
    },
    xlabel={Number of votes},
    xmin=0,
    xmax=16,
    xtick={0,1,2,3,4,5,6,7,8,9,10,11,12,13,14,15,16},
    ytick=data,
    yticklabels={\textit{uss},\textit{stem sep.},\textit{search},\textit{enhancement},\textit{reverb},\textit{bgm},\textit{upmixing}},
    xticklabels={0,,2,,4,,6,,8,,10,,12,,14,,16},
    point meta=explicit symbolic,
]

\addplot[draw=gray,fill=gray!15] coordinates {
    (13,7)[72.2\%]  
    (12,6)[66.7\%]  
    (9,5)[50.0\%]  
    (8,4)[44.4\%]  
    (8,3)[44.4\%]  
    (4,2)[22.2\%]  
    (2,1)[11.1\%]  
};

\end{axis}
\end{tikzpicture}%
\captionsetup{width=0.8\textwidth}
\caption{
    Answers of a multiple-choice question. \vspace{5pt} \\
    \textbf{Q}: Except for generative audio AI, which technology do you think would be useful? \vspace{3pt} \\
    \textbf{A1}: \textit{uss} - Universal source separation with text condition \\
    \textbf{A2}: \textit{stem sep.} - Automatic separation of stems from mixed or mastered tracks\\
    \textbf{A3}: \textit{search} - Simplified and efficient search algorithms or visualization methods \\
    \textbf{A4}: \textit{enhancement} - Audio enhancement to improve audio quality, such as sample rate or fidelity \\
    \textbf{A5}: \textit{reverb} - De-reverberation or room-impulse response estimation \\
    \textbf{A6}: \textit{bgm} - Automatic rearrangement for background music \\
    \textbf{A7}: \textit{upmixing} - Automatic upmixing (e.g. mono to 5.1ch audio) \\
}
\end{figure}

\begin{figure}[H]
\centering
\begin{tikzpicture}[trim axis left]
\begin{axis} [
    xbar,
    width=0.6\textwidth,
    y=17pt,
    tick align=outside,
    enlargelimits=0,
    enlarge y limits={abs=0.6},
    nodes near coords,
    nodes near coords align=horizontal,
    nodes near coords style={
        anchor=west,
    },
    xlabel={Number of votes},
    xmin=0,
    xmax=12,
    xtick={0,1,2,3,4,5,6,7,8,9,10,11,12,13},
    ytick=data,
    yticklabels={\textit{stereo-wet},\textit{stereo-dry},\textit{mono-wet},\textit{mono-dry},\textit{no matter}},
    xticklabels={0,,2,,4,,6,,8,,10,,12,,},
    point meta=explicit symbolic,
]

\addplot[draw=gray,fill=gray!15] coordinates {
    (4,1)[22.2\%]  
    (9,2)[50.0\%]  
    (1,3)[5.6\%]  
    (10,4)[55.6\%]  
    (5,5)[27.8\%]  
};

\end{axis}
\end{tikzpicture}%
\captionsetup{width=0.8\textwidth}
\caption{
    Answers of a multiple-choice question. \vspace{5pt} \\
    \textbf{Q} What audio type do you prefer for output? \vspace{3pt} \\
    \textbf{A1}: \textit{no matter} - All possible types \\
    \textbf{A2}: \textit{mono-dry} - 1-channel audio signal without any reverb \\
    \textbf{A3}: \textit{mono-wet} - 1-channel audio signal with proper reverb \\
    \textbf{A4}: \textit{stereo-dry} - 2-channel audio signal without any reverb \\
    \textbf{A5}: \textit{stereo-wet} - 2-channel audio signal with proper reverb \\
}
\end{figure}

\section{Experiment Results for Section \ref{sec:challenges}}

\begin{figure}[H]
    \centering
    \includegraphics[width=0.96\linewidth]{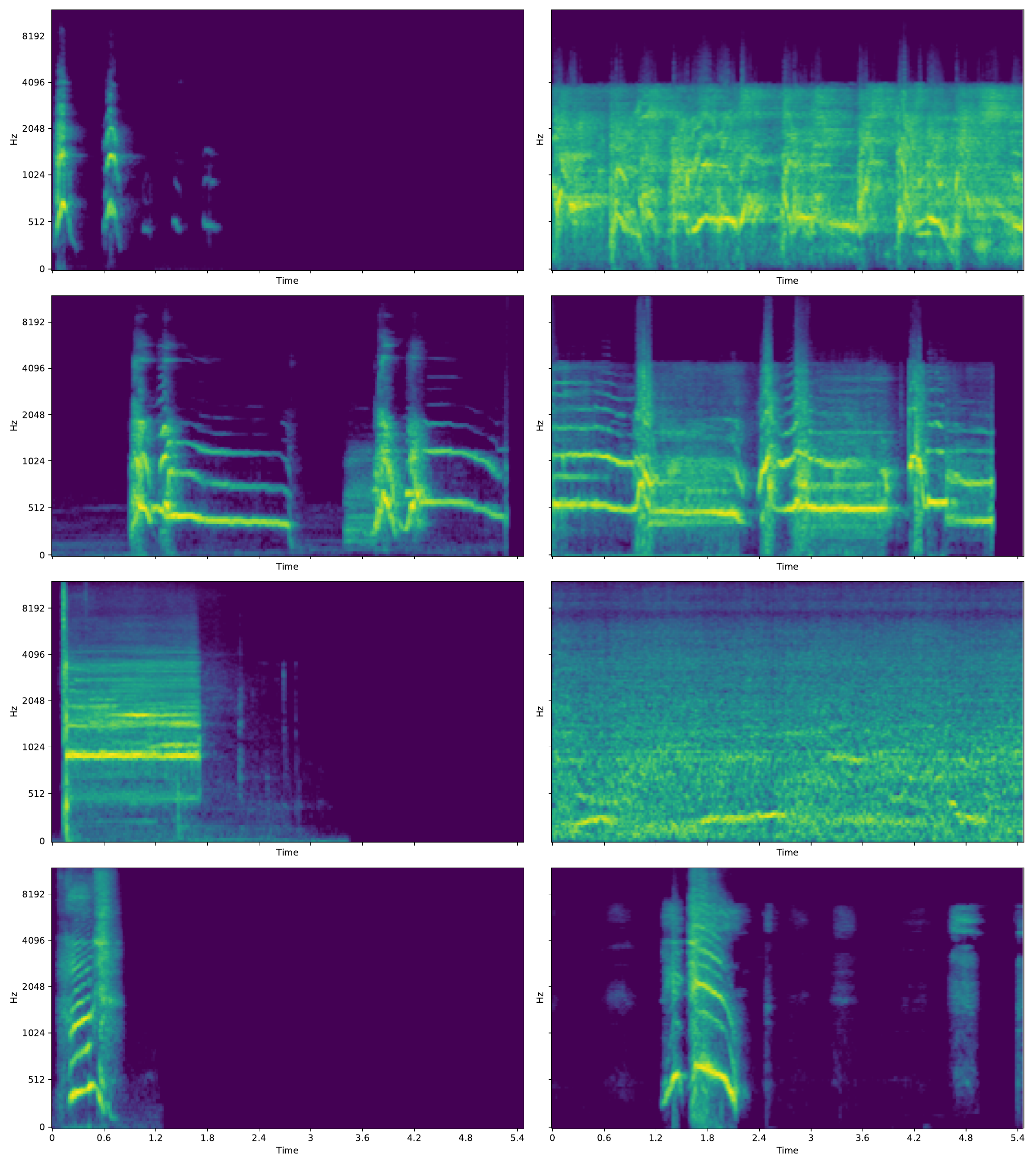}
\caption{Mel-spectrograms of the generated audio samples with different dataset labels -- \textit{left}: with ``clean" label, and \textit{right}: with ``noisy" label. Samples in the same row are generated with the same prompt. Prompts used to generate samples in each row are as follows; (i) ``small dog bark", (ii) ``dog howling", (iii) ``the sound of starting a car engine", and (iv) ``male sneeze".
With the \textit{clean} dataset label, the model generates high quality signals with less background noise and more high frequency components. Conversely, generated samples show low quality results when the model is conditioned with the \textit{noisy} label. Specifically, most of noisy-labeled signals show underlying noise or interference and have limited bandwidth for some samples.}
\end{figure}

%% file: main.bbl
\begin{thebibliography}{31}
\providecommand{\natexlab}[1]{#1}
\providecommand{\url}[1]{\texttt{#1}}
\expandafter\ifx\csname urlstyle\endcsname\relax
  \providecommand{\doi}[1]{doi: #1}\else
  \providecommand{\doi}{doi: \begingroup \urlstyle{rm}\Url}\fi

\bibitem[Chen et~al.(2020)Chen, Xie, Vedaldi, and Zisserman]{chen2020vggsound}
Chen, H., Xie, W., Vedaldi, A., and Zisserman, A.
\newblock Vggsound: A large-scale audio-visual dataset.
\newblock In \emph{ICASSP 2020-2020 IEEE International Conference on Acoustics,
  Speech and Signal Processing (ICASSP)}, pp.\  721--725. IEEE, 2020.

\bibitem[Choi et~al.(2022)Choi, Oh, Kang, and McFee]{choi2022proposal}
Choi, K., Oh, S., Kang, M., and McFee, B.
\newblock A proposal for foley sound synthesis challenge, 2022.

\bibitem[Choi et~al.(2023)Choi, Im, Heller, McFee, Imoto, Okamoto, Lagrange,
  and Takamichi]{choi2023foley}
Choi, K., Im, J., Heller, L., McFee, B., Imoto, K., Okamoto, Y., Lagrange, M.,
  and Takamichi, S.
\newblock Foley sound synthesis at the dcase 2023 challenge, 2023.

\bibitem[Chowdhery et~al.(2022)Chowdhery, Narang, Devlin, Bosma, Mishra,
  Roberts, Barham, Chung, Sutton, Gehrmann, et~al.]{chowdhery2022palm}
Chowdhery, A., Narang, S., Devlin, J., Bosma, M., Mishra, G., Roberts, A.,
  Barham, P., Chung, H.~W., Sutton, C., Gehrmann, S., et~al.
\newblock Palm: Scaling language modeling with pathways.
\newblock \emph{arXiv preprint arXiv:2204.02311}, 2022.

\bibitem[Chowning(1973)]{chowning1973synthesis}
Chowning, J.~M.
\newblock The synthesis of complex audio spectra by means of frequency
  modulation.
\newblock \emph{Journal of the audio engineering society}, 21\penalty0
  (7):\penalty0 526--534, 1973.

\bibitem[Cui et~al.(2023)Cui, Zhao, Ren, Liu, Huang, Chen, Wang, Huai, and
  Wu]{cui2023varietysound}
Cui, C., Zhao, Z., Ren, Y., Liu, J., Huang, R., Chen, F., Wang, Z., Huai, B.,
  and Wu, F.
\newblock Varietysound: Timbre-controllable video to sound generation via
  unsupervised information disentanglement.
\newblock In \emph{ICASSP 2023-2023 IEEE International Conference on Acoustics,
  Speech and Signal Processing (ICASSP)}, pp.\  1--5. IEEE, 2023.

\bibitem[Drossos et~al.(2020)Drossos, Lipping, and Virtanen]{drossos2020clotho}
Drossos, K., Lipping, S., and Virtanen, T.
\newblock Clotho: An audio captioning dataset.
\newblock In \emph{ICASSP 2020-2020 IEEE International Conference on Acoustics,
  Speech and Signal Processing (ICASSP)}, pp.\  736--740. IEEE, 2020.

\bibitem[Dudley(1955)]{dudley1955fundamentals}
Dudley, H.
\newblock Fundamentals of speech synthesis.
\newblock \emph{Journal of the Audio Engineering Society}, 3\penalty0
  (4):\penalty0 170--185, 1955.

\bibitem[Fonseca et~al.(2021)Fonseca, Favory, Pons, Font, and
  Serra]{fonseca2021fsd50k}
Fonseca, E., Favory, X., Pons, J., Font, F., and Serra, X.
\newblock Fsd50k: an open dataset of human-labeled sound events.
\newblock \emph{IEEE/ACM Transactions on Audio, Speech, and Language
  Processing}, 30:\penalty0 829--852, 2021.

\bibitem[Font et~al.(2013)Font, Roma, and Serra]{font2013freesound}
Font, F., Roma, G., and Serra, X.
\newblock Freesound technical demo.
\newblock In \emph{Proceedings of the 21st ACM international conference on
  Multimedia}, pp.\  411--412, 2013.

\bibitem[Gemmeke et~al.(2017)Gemmeke, Ellis, Freedman, Jansen, Lawrence, Moore,
  Plakal, and Ritter]{gemmeke2017audio}
Gemmeke, J.~F., Ellis, D.~P., Freedman, D., Jansen, A., Lawrence, W., Moore,
  R.~C., Plakal, M., and Ritter, M.
\newblock Audio set: An ontology and human-labeled dataset for audio events.
\newblock In \emph{2017 IEEE international conference on acoustics, speech and
  signal processing (ICASSP)}, pp.\  776--780. IEEE, 2017.

\bibitem[Ghose \& Prevost(2020)Ghose and Prevost]{ghose2020autofoley}
Ghose, S. and Prevost, J.~J.
\newblock Autofoley: Artificial synthesis of synchronized sound tracks for
  silent videos with deep learning.
\newblock \emph{IEEE Transactions on Multimedia}, 23:\penalty0 1895--1907,
  2020.

\bibitem[Ghose \& Prevost(2022)Ghose and Prevost]{ghose2022foleygan}
Ghose, S. and Prevost, J.~J.
\newblock Foleygan: Visually guided generative adversarial network-based
  synchronous sound generation in silent videos.
\newblock \emph{IEEE Transactions on Multimedia}, 2022.

\bibitem[Huang et~al.(2023)Huang, Huang, Yang, Ren, Liu, Li, Ye, Liu, Yin, and
  Zhao]{huang2023make}
Huang, R., Huang, J., Yang, D., Ren, Y., Liu, L., Li, M., Ye, Z., Liu, J., Yin,
  X., and Zhao, Z.
\newblock Make-an-audio: Text-to-audio generation with prompt-enhanced
  diffusion models.
\newblock \emph{arXiv preprint arXiv:2301.12661}, 2023.

\bibitem[Kang et~al.(2023{\natexlab{a}})Kang, Oh, Moon, Lee, and
  Chon]{ChonGLI2023}
Kang, M., Oh, S., Moon, H., Lee, K., and Chon, B.~S.
\newblock Fall-e: Gaudio foley synthesis system.
\newblock Technical report, Gaudio Lab, Inc., Seoul, South Kore, June
  2023{\natexlab{a}}.

\bibitem[Kang et~al.(2023{\natexlab{b}})Kang, Oh, Moon, Lee, and
  Chon]{kang2023falle}
Kang, M., Oh, S., Moon, H., Lee, K., and Chon, B.~S.
\newblock Fall-e: A foley sound synthesis model and strategies,
  2023{\natexlab{b}}.

\bibitem[Kim et~al.(2019)Kim, Kim, Lee, and Kim]{kim-etal-2019-audiocaps}
Kim, C.~D., Kim, B., Lee, H., and Kim, G.
\newblock {A}udio{C}aps: Generating captions for audios in the wild.
\newblock In \emph{Proceedings of the 2019 Conference of the North {A}merican
  Chapter of the Association for Computational Linguistics: Human Language
  Technologies, Volume 1 (Long and Short Papers)}, pp.\  119--132, Minneapolis,
  Minnesota, June 2019. Association for Computational Linguistics.
\newblock \doi{10.18653/v1/N19-1011}.
\newblock URL \url{https://aclanthology.org/N19-1011}.

\bibitem[Kreuk et~al.(2022)Kreuk, Synnaeve, Polyak, Singer, D{\'e}fossez,
  Copet, Parikh, Taigman, and Adi]{kreuk2022audiogen}
Kreuk, F., Synnaeve, G., Polyak, A., Singer, U., D{\'e}fossez, A., Copet, J.,
  Parikh, D., Taigman, Y., and Adi, Y.
\newblock Audiogen: Textually guided audio generation.
\newblock \emph{arXiv preprint arXiv:2209.15352}, 2022.

\bibitem[Liu et~al.(2023)Liu, Chen, Yuan, Mei, Liu, Mandic, Wang, and
  Plumbley]{liu2023audioldm}
Liu, H., Chen, Z., Yuan, Y., Mei, X., Liu, X., Mandic, D., Wang, W., and
  Plumbley, M.~D.
\newblock Audioldm: Text-to-audio generation with latent diffusion models.
\newblock \emph{arXiv preprint arXiv:2301.12503}, 2023.

\bibitem[OpenAI(2023)]{openai2023gpt4}
OpenAI.
\newblock Gpt-4 technical report, 2023.

\bibitem[Piczak(2015)]{piczak2015esc}
Piczak, K.~J.
\newblock Esc: Dataset for environmental sound classification.
\newblock In \emph{Proceedings of the 23rd ACM international conference on
  Multimedia}, pp.\  1015--1018, 2015.

\bibitem[Ramesh et~al.(2021)Ramesh, Pavlov, Goh, Gray, Voss, Radford, Chen, and
  Sutskever]{ramesh2021zero}
Ramesh, A., Pavlov, M., Goh, G., Gray, S., Voss, C., Radford, A., Chen, M., and
  Sutskever, I.
\newblock Zero-shot text-to-image generation.
\newblock In \emph{International Conference on Machine Learning}, pp.\
  8821--8831. PMLR, 2021.

\bibitem[Ramesh et~al.(2022)Ramesh, Dhariwal, Nichol, Chu, and
  Chen]{ramesh2022hierarchical}
Ramesh, A., Dhariwal, P., Nichol, A., Chu, C., and Chen, M.
\newblock Hierarchical text-conditional image generation with clip latents.
\newblock \emph{arXiv preprint arXiv:2204.06125}, 2022.

\bibitem[Rombach et~al.(2022)Rombach, Blattmann, Lorenz, Esser, and
  Ommer]{rombach2022high}
Rombach, R., Blattmann, A., Lorenz, D., Esser, P., and Ommer, B.
\newblock High-resolution image synthesis with latent diffusion models.
\newblock In \emph{Proceedings of the IEEE/CVF Conference on Computer Vision
  and Pattern Recognition}, pp.\  10684--10695, 2022.

\bibitem[Saharia et~al.(2022)Saharia, Chan, Saxena, Li, Whang, Denton,
  Ghasemipour, Gontijo~Lopes, Karagol~Ayan, Salimans,
  et~al.]{saharia2022photorealistic}
Saharia, C., Chan, W., Saxena, S., Li, L., Whang, J., Denton, E.~L.,
  Ghasemipour, K., Gontijo~Lopes, R., Karagol~Ayan, B., Salimans, T., et~al.
\newblock Photorealistic text-to-image diffusion models with deep language
  understanding.
\newblock \emph{Advances in Neural Information Processing Systems},
  35:\penalty0 36479--36494, 2022.

\bibitem[Salamon et~al.(2014)Salamon, Jacoby, and Bello]{salamon2014dataset}
Salamon, J., Jacoby, C., and Bello, J.~P.
\newblock A dataset and taxonomy for urban sound research.
\newblock In \emph{Proceedings of the 22nd ACM international conference on
  Multimedia}, pp.\  1041--1044, 2014.

\bibitem[Schuhmann et~al.(2021)Schuhmann, Vencu, Beaumont, Kaczmarczyk, Mullis,
  Katta, Coombes, Jitsev, and Komatsuzaki]{schuhmann2021laion}
Schuhmann, C., Vencu, R., Beaumont, R., Kaczmarczyk, R., Mullis, C., Katta, A.,
  Coombes, T., Jitsev, J., and Komatsuzaki, A.
\newblock Laion-400m: Open dataset of clip-filtered 400 million image-text
  pairs.
\newblock \emph{arXiv preprint arXiv:2111.02114}, 2021.

\bibitem[Touvron et~al.(2023)Touvron, Lavril, Izacard, Martinet, Lachaux,
  Lacroix, Rozi{\`e}re, Goyal, Hambro, Azhar, et~al.]{touvron2023llama}
Touvron, H., Lavril, T., Izacard, G., Martinet, X., Lachaux, M.-A., Lacroix,
  T., Rozi{\`e}re, B., Goyal, N., Hambro, E., Azhar, F., et~al.
\newblock Llama: Open and efficient foundation language models.
\newblock \emph{arXiv preprint arXiv:2302.13971}, 2023.

\bibitem[Wu et~al.(2023)Wu, Chen, Zhang, Hui, Berg-Kirkpatrick, and
  Dubnov]{wu2023large}
Wu, Y., Chen, K., Zhang, T., Hui, Y., Berg-Kirkpatrick, T., and Dubnov, S.
\newblock Large-scale contrastive language-audio pretraining with feature
  fusion and keyword-to-caption augmentation.
\newblock In \emph{ICASSP 2023-2023 IEEE International Conference on Acoustics,
  Speech and Signal Processing (ICASSP)}, pp.\  1--5. IEEE, 2023.

\bibitem[Yang et~al.(2023)Yang, Yu, Wang, Wang, Weng, Zou, and
  Yu]{yang2023diffsound}
Yang, D., Yu, J., Wang, H., Wang, W., Weng, C., Zou, Y., and Yu, D.
\newblock Diffsound: Discrete diffusion model for text-to-sound generation.
\newblock \emph{IEEE/ACM Transactions on Audio, Speech, and Language
  Processing}, 2023.

\bibitem[Zhou et~al.(2018)Zhou, Wang, Fang, Bui, and Berg]{zhou2018visual}
Zhou, Y., Wang, Z., Fang, C., Bui, T., and Berg, T.~L.
\newblock Visual to sound: Generating natural sound for videos in the wild.
\newblock In \emph{Proceedings of the IEEE conference on computer vision and
  pattern recognition}, pp.\  3550--3558, 2018.

\end{thebibliography}
